\documentclass{elsart}

\newcommand{\ms}{$M_{\odot}$}
\newcommand{\msb}{$M_{\odot}$~}
\newcommand{\al}{$^{26}$Al}

\usepackage{natbib}
 \usepackage{graphicx}

\usepackage{amssymb}

 \usepackage{lineno}

\begin{document}

\begin{frontmatter}



\title{$^{26}$Al production from
magnetically induced extramixing in AGB Stars}


\author{S. Palmerini, M. Busso}
{\address{Dipartimento di Fisica, Universit$\grave{a}$ degli Studi
di Perugia and
\\INFN sezione di Perugia, via Pascoli, 06123 Perugia, Italy} }

\begin{abstract}
We discuss nucleosynthesis results obtained following the recent
suggestion that extramixing phenomena in red giants might be
driven by magnetic buoyancy. We explore for this model the
production of the short-lived radioactive isotope $^{26}$Al and of
stable light nuclei, considering both the case of the general
buoyancy of flux tubes and that of the intermittent release of
magnetized unstable structures. We show that abundant $^{26}$Al
can be produced, up to, and above, the highest levels measured in
presolar grains. This level would be also sufficient to explain
the early solar system $^{26}$Al as coming from a nearby AGB star
of low mass. The case of fast-moving instabilities is the most
efficient, reaching almost the same effectiveness as hot bottom
burning (HBB).
\end{abstract}


\end{frontmatter}

\section{Introduction}

Previous studies have shown that AGB stars are a site where
several short-lived radioactivities identified in early solar
system (ESS) materials are produced \citep{w95,bgw}. In
particular, the possibility that one such star polluted the early
solar nebula, introducing the record of nuclei like $^{26}$Al,
$^{41}$Ca, $^{60}$Fe, $^{107}$Pd, $^{135}$Cs, was explored by
\citet{w94}. They found that the polluting star must also
experience extra-mixing phenomena, carrying to the envelope
partially H-processed materials rich in \al~ \citep{wbgn}. These
extended mixing episodes were also shown to account for the record
of C and O isotopes in presolar grains that also show \al~
\citep{nol}. In the mentioned models circulation was treated
parametrically, and the physical mechanism was left unspecified.

Among the processes invoked to explain extended mixing, rotation
played a dominant role \citep{D}, as it can induce shear
instabilities and meridian circulation. However, attempts at
considering rotational mixing in complete stellar models failed to
reproduce the observed abundance changes \citep{pal}. This fact
promoted a reappraisal of other forms of matter transport, like
the so-called ''tachohaline'' mixing induced by the decrease in
the molecular weight created by $^3$He combustion
\citep{eggl,chaz}. The efficiency of such a mechanism, very
important in other contexts, seems however to be in question in
the environment we refer to, i.e. in AGB stars of about 2 \msb
\citep{can}.

Guided by the existing models of magnetic buoyancy \citep{parker}
and of other magnetic instabilities \citep{spru1,spru9,egg},
\citet{wbc} [hereafter BWNC] recently proposed that matter transport
in red giants be maintained by a magnetic dynamo, creating a system
of buoyant toroidal flux tubes. In this paper we shall explore in a
preliminary way some of the nucleosynthesis implications of such an
idea. In particular, we analyze the synthesis of $^{26}$Al in AGB
stars, implementing in a simplified way a mass circulation based on
BWNC's formalism.

\section{The Model and the Numerical technique}

From the original poloidal field of a rotating star, a toroidal
field of similar strength can be generated by the dynamo mechanism
\citep{parker}. In physical conditions occurring above the H-burning
shell of an evolved star several perturbations can affect a
cylindrical or toroidal flux tube, and they can evolve into
instabilities \citep{spru9}. These are opposed by the magnetic field
aligned with the cylinder, which acts as a stabilizing tension. The
complex dynamical behavior of such a tube includes expansion (due to
buoyancy), unstable oscillations (due to Alfv\'en waves, leading to
the development of $\Omega$-shaped loops) and torsion due to the
Coriolis force \citep{spru2}. The absence of X-ray emission from
surface magnetic activity in AGB stars tells us that these
magnetized structures, if they do exist, must be disrupted in the
convective layers, thus depositing there their chemical
peculiarities. BWNC showed that efficient buoyancy of magnetized
regions requires that they achieve very high values of the magnetic
fields, up to a few MG.

\begin{figure*}[t!!]
\centering \resizebox{\textwidth}{!}
{\includegraphics[height=5cm]{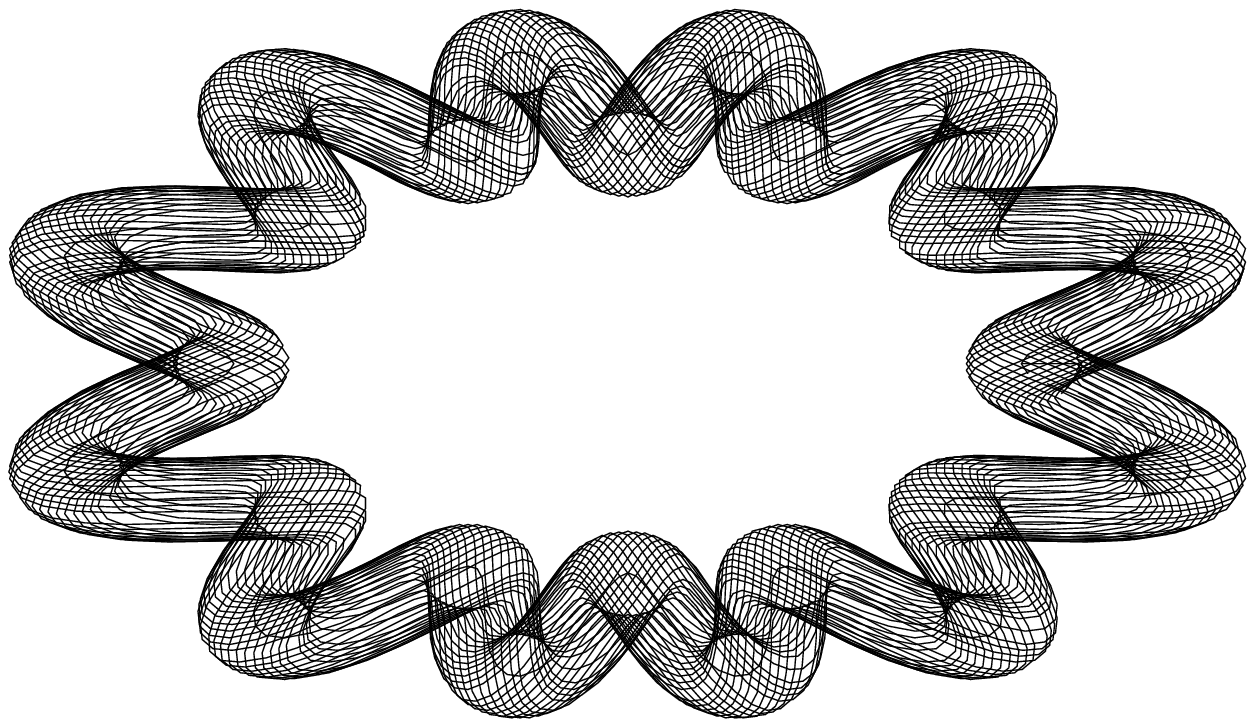}
\includegraphics[height=5cm]{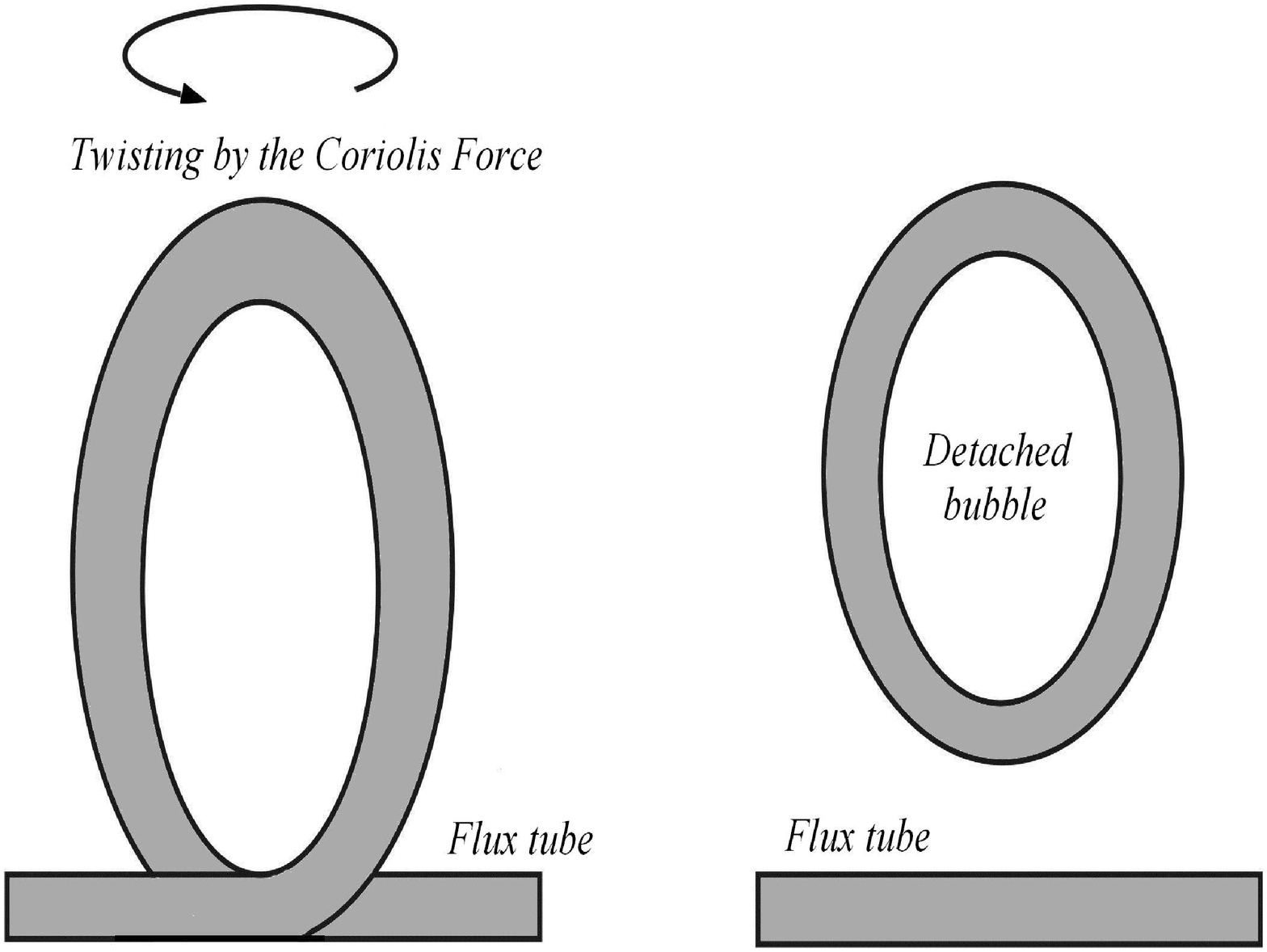}}
\caption{Left. A buoyant flux tube, laying in the equatorial plane,
with Alfv\'en wave oscillations. If the field is strong enough (MG)
it floats to the envelope, depositing there its composition. Right.
Detachment of a magnetized structure due to the tilting and
reconnection of an unstable $\Omega$-shaped loop. The released
bubbles move rapidly to the envelope, carrying material synthesized
close to the H-shell}\label{pbf1}
\end{figure*}

It is difficult for a star to maintain very strong fields buried
in its interior: if the {\it average} field is strong, than it is
ejected in a short interval of time and a whole toroidal flux tube
can be assumed to float to the envelope (see Figure \ref{pbf1},
left panel). In this case, maintaining the dynamo for subsequent
episodes of magnetic activity depends on the possibility that
energy is resupplied from the convective envelope to the interior
and the differential rotation is re-established \citep{nb}. If
this occurs, than flux tubes emerge, break into the envelope and
force local matter to be pushed down in a downflow. This probably
occurs over a wide area, and reaches down to regions where
p-captures are active. The upward velocity $v_u$ of the buoyancy
can be taken from BWNC. Assuming local mass conservation in each
shell crossed by the upward and downward motions, the downward
velocity would be such that $v_d f_d$ = $v_u f_u$, where $f_d$ and
$f_u$ are the fractional areas occupied by the upflow and by the
downflow, respectively. The dimensions of a flux tube as discussed
by BWNC require that $f_d/f_u$ be very large ($>$200) even in the
case where the activity phenomena related to toroidal fields are
constrained within a limited range in latitude (in the Sun, this
is normally $\pm$ 30-50$^o$, corresponding to $f_d = 0.5-0.75$).
In such conditions, H burning has time to occur only in the
downward movement, due to the lower velocity. When a new magnetic
flux tube forms, everything is repeated: the net effect is a sort
of matter circulation. We shall refer to this scheme as Case 1. As
discussed in BWNC, the chemical changes to be explained require
that, in the process, the envelope is mixed with a roughly equal
mass of partially H-burned material.

Alternatively, magnetic instabilities might be released
intermittently. This corresponds, for example, to the scheme of
Figure \ref{pbf1}, right panel, in which a kink of an Alfv\'en
oscillation evolves into an $\Omega$-shaped instability of high
internal fields. It can then be tilted by the Coriolis force, and
detached from the rest of the flux tube as an independent bubble. We
shall refer to this scheme as Case 2. We shall present here
preliminary results for both cases.

\section{Effects of magnetic extra-mixing on $^{26}$Al and other isotopes}

We use the model for a star of 2\msb and solar metallicity,
previously calculated with the FRANEC code using a limited
reaction network for energy production. A mass loss rate as
described by the parametrization by \citet{rei}, with $\eta = $
0.5 was used. On this basis we recomputed the full nucleosynthesis
in the radiative layers above the H shell and followed the mixing
of processed materials into the convective envelope. For Case 1,
this was done in a way similar to that of \citet{nol}, but
adopting the upward velocity given in BWNC and taking $f_d$ =
0.75. For Case 2, we assumed instead that every 50 years the
phenomenon of Figure \ref{pbf1} (right panel) occurs, carrying to
the surface H-processed material. The period of 50 years was
chosen because it is comparable to the magnetic cycles in active
stars and to the duration of magnetic bursts in evolved stars
found by \citet{nord}.

In Case 1 we assume a maximum temperature for the circulation of
$Log T_P = Log T_H -0.1$. This condition was taken from
\citet{nol}, who showed that for higher temperatures excessive
energy production would occur. In the second case, due to the long
duration of the burning at $T_P$ in between two episodes of bubble
release, we had to reduce $T_P$ not to disturb the stellar
structure ($\Delta Log T_P = 0.15$). In several runs we explored
the range of mass circulation from $\dot M$=$10^{-7}$ to  $\dot
M$=$10^{-5}$ \ms$/yr$, following the indications by \citet{nol}
that this is the interval in which the isotopic shifts observed in
presolar grains of AGB origin can be explained. We notice that,
although the value of $\dot M$ is not self-consistently obtained,
but rather adopted from the quoted work, there are specific
relations linking the mass circulation to the magnetic field $B$
(at least in Case 1). Indeed, for flux conservation, calling $a$
the radius of the flux tube, $\dot M \propto B a^{2}$. On the
other hand, the buoyancy velocity $v$ is related to the field by
the relation $v \propto B$ \citep[see][Eq.4]{wbc}, therefore
$\dot{M}\propto B \times a^{\frac{3}{2}}$.
\\Tube dimensions were assumed from the solar case;
then increasing $B$ is required to obtain higher $\dot M$ values.
Similarly, lower mass circulations would imply lower fields, as
indeed found by BWNC for RGB stars. While small vales of the mass
circulation are certainly possible, it seems unlikely that a much
faster transport (say, by a factor 100) can be magnetically
supported, unless the tube dimensions are very different from the
solar case. Indeed, the field values obtained by BWNC for $\dot M
\sim 10^{-6}M_{\odot}/yr$ are already quite high. In the scheme of
Case 2, the analogy with solar magnetic tubes is less important,
because the achieved envelope abundances are essentially controlled
only by the long burning periods (tens of years) at $T_P$. In this
case, magnetic buoyancy simply transports in discrete bubbles matter
synthesized by the H shell, but continuously resupplying the fuel
with material descending from the envelope (by mass conservation).
High $B$ values are required only in the buoyant bubbles, not in the
buried magnetic structures, and a faster mass transport is a priori
possible.

In general, computing extra-mixing requires describing nuclear
physics phenomena coupled with dynamics. This can be obtained by
solving a system of differential equation of the type:
\begin{equation}
\label{eq2}
\frac{dN_i}{dt}=\frac{\partial{N_i}}{\partial{t}}+\frac{\partial{N_i}}
{\partial{M}}\frac{\partial{M}}{\partial{R}}\frac{\partial{R}}{\partial{t}}
\end{equation}
where the partial time derivative due to nucleosynthesis is:
\begin{equation}
\label{eq1}
\frac{\partial{N_i}}{\partial{t}}=-N_{p}N_{i}\lambda_{i,p}+N_{p}N_{i-1}\lambda_{i-1,p}-N_{i}\lambda_{d}+N_{i'}\lambda_{d}
\end{equation}
Here the parameters $\lambda$ are the reaction rates, which we
take from the NACRE compilation \citep{NACRE}. In practical cases,
the phenomenon might become simpler than that. For example, in the
fast upward flux the time scale for motion is smaller than the one
for burning, and no nucleosynthesis occurs. On the contrary, in a
sufficiently slow descending flux everything depends only on the
path integral of reaction rates, as in previous extramixing
calculations by \citet{w95}

In our work we first recomputed the ''static'' H burning through the
radiative region for the whole AGB phase, with an extended network
going from H to $^{32}$S. Then we applied separately our mixing
schemes for Cases 1 and 2.

\begin{figure*}[t!!]
\centering \resizebox{\textwidth}{!}
{\includegraphics[height=7cm]{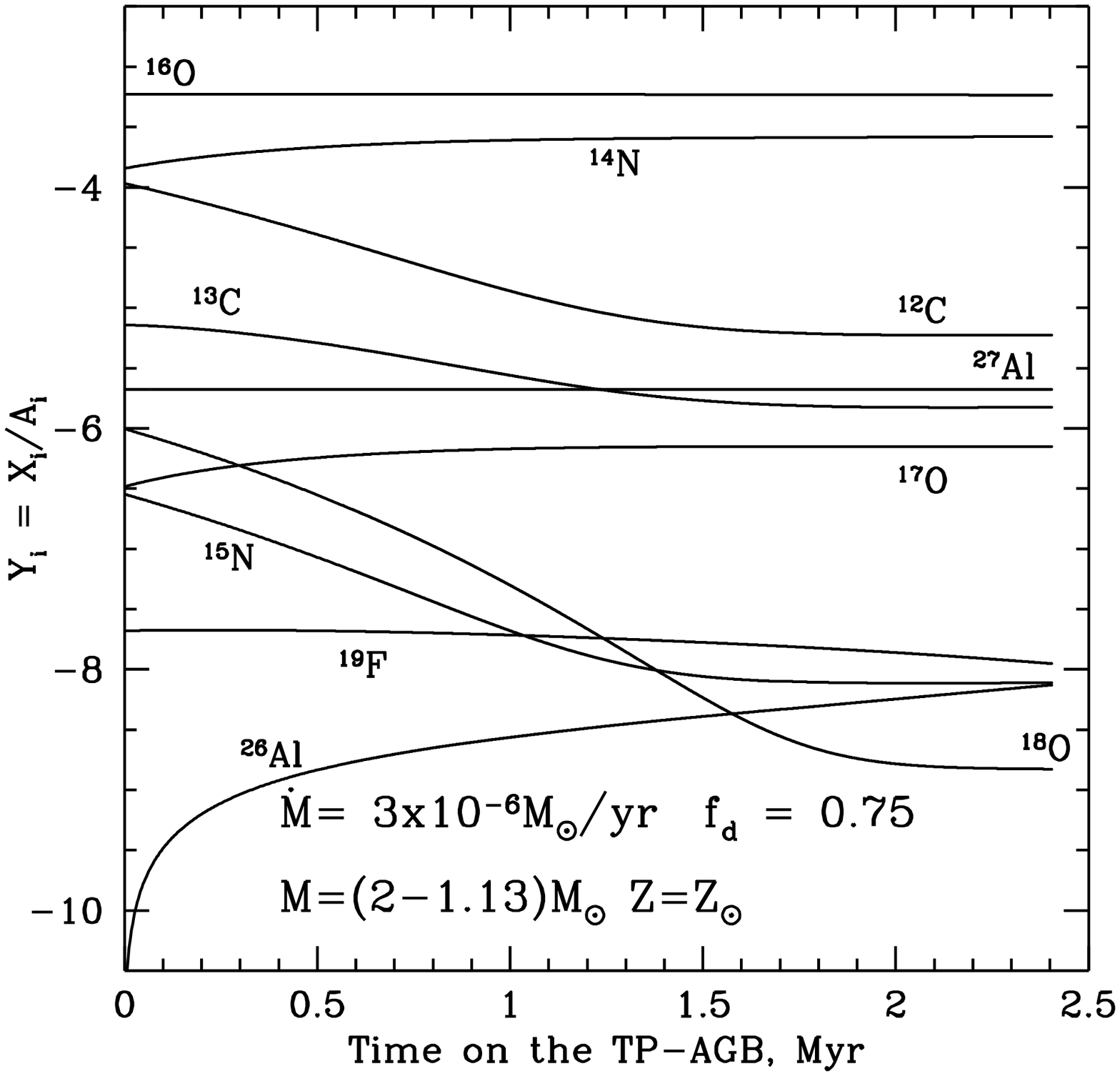}
\includegraphics[height=7.1cm]{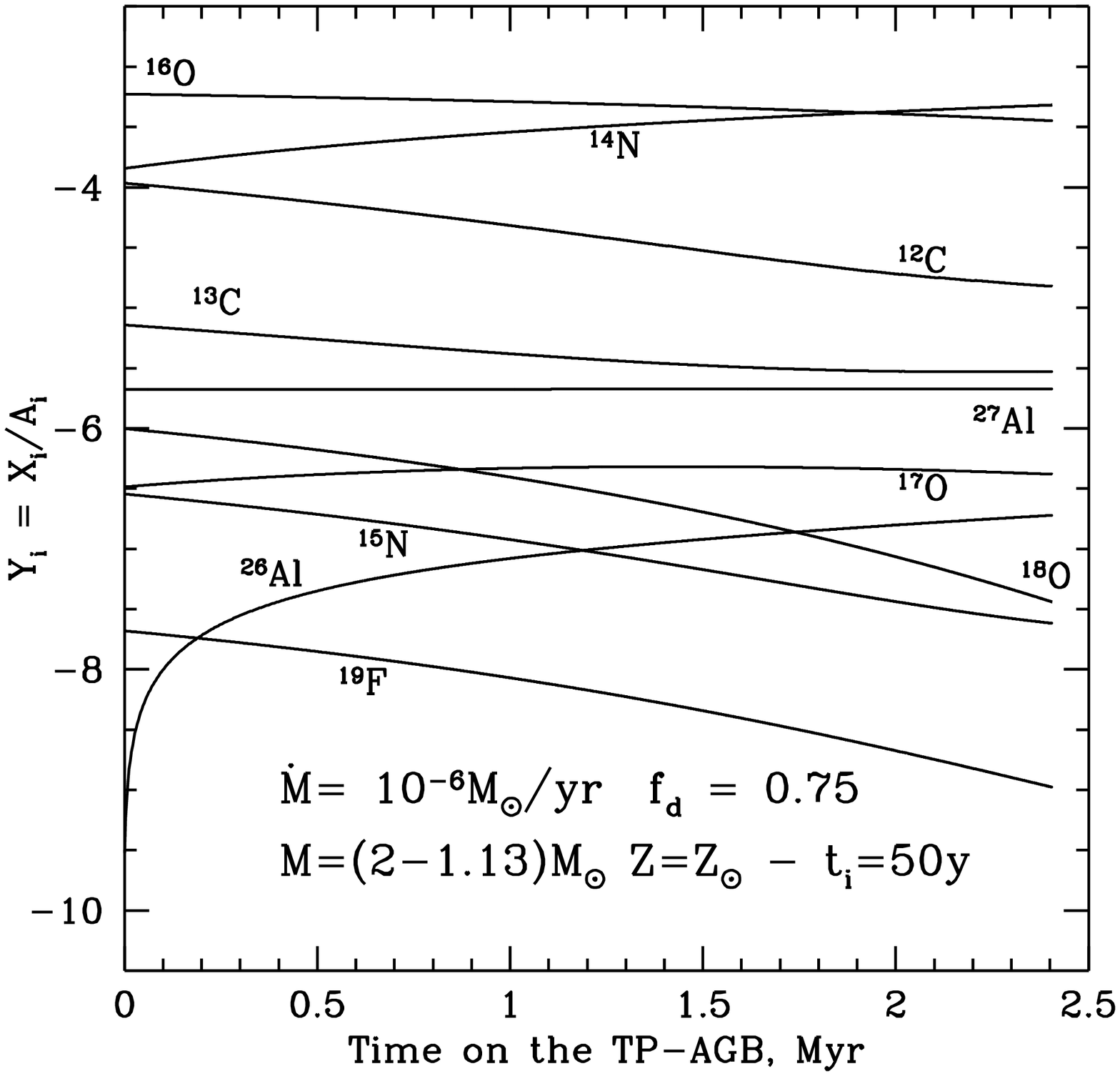}}
\caption{Left: envelope abundance evolution along the TP-AGB phase
for nuclei affected by CBP, in a star of $2 M_{\odot}$ and
$Z_{\odot}$, in Case 1 (circulation). Right:  envelope abundances
in Case 2 (detached bubbles). Here $t_i$ is the the time spent
between the release of two successive bubbles} \label{pbf2}
\end{figure*}

Figure \ref{pbf2} shows preliminary results thus obtained, in terms
of abundances of affected isotopes in the envelope, as a function of
the time spent on the thermally-pulsing AGB stage. The changes in
abundances introduced by the third dredge-up have for the moment
been neglected.

Preliminary conclusions drawn from Figure \ref{pbf2} can be
summarized as follows:
\begin{itemize}
\item{Case 1 (Figure \ref{pbf2}, left panel) gives results very
similar to what was previously obtained in parameterized
extramixing (or cool bottom processing, CBP, as this phenomenon is
sometimes called) by \citet{nol}. $^{26}$Al is produced at a
moderate efficiency, reaching final abundance ratios in the
envelope in the range \al/$^{27}$Al = 10$^{-3}$ to 10$^{-2}$
(depending on the choice of parameters). This corresponds to
values found by \citet{wbgn}. $^{16}$O is virtually untouched
(i.e. we have only CN cycling), while heavier oxygen isotopes are
largely modified, covering the region where presolar grain data
lay}.

\item{Case 2 (Figure \ref{pbf2}, right panel) shows a somewhat
different burning process. The abundances of nuclei modified at
high $T$ (like $^{16}$O) are more affected, and
$^{26}$Al/$^{27}$Al reaches values close to 0.1. By contrast,
fragile nuclei like $^{18}$O and $^{15}$N, already burning at low
T, are less affected than in Case 1, where they have more time to
be consumed. The high abundance of $^{26}$Al found makes this case
interesting for the possibility that the encounter of an AGB star
be at the origin of the early solar system concentration of
$^{26}$Al}.
\end{itemize}
\section{Acknowledgement}
This work was supported by the Italian Ministry of Research
(Contract PRIN-2006-022731). Without the close collaboration, the
lively discussions, the friendly advise and the continuous support
by G.J. Wasserburg and K.M. Nollett this work would have been
impossible.




\end{document}